\documentclass[journal]{IEEEtran}
\usepackage{amsmath}
\usepackage{amssymb}
\usepackage{latexsym}
\usepackage{multirow}
\usepackage{epsfig}
\usepackage{graphics}

\newcommand{\eqdef}{\stackrel{\triangle}{=}}
\newcommand{\beq}{\begin{equation}}
\newcommand{\enq}{\end{equation}}
\newcommand{\ben}{\begin{eqnarray}}
\newcommand{\enn}{\end{eqnarray}}

\begin{document}

\title{Optimal Upward Scaling of Minimum-TSC\\Binary Signature
Sets}

\author{Lili~Wei,~\IEEEmembership{Member,~IEEE}, and Wen Chen,~\IEEEmembership{Senior Member,~IEEE}
\thanks{Manuscript received July 13, 2011; revised October 12, 2011, accepted November 27th, 2011. The
associate editor coordinating the review of this paper and approving
it for publication was S. Kotsopoulos.}
\thanks{L. Wei, W. Chen are
with the Department of Electronic Engineering, Shanghai Jiao Tong
University, Shanghai, China, and SKL for ISN, Xidian University,
China (e-mail: lili.wei@sjtu.edu.cn; wenchen@sjtu.edu.cn).}
\thanks{This work is supported by NSF China \#60972031, by national 973
project \#2012CB316106 and \#2009CB824900, by national huge special
project \#2012ZX03004004, by national key laboratory project
\#ISN11-01, by Huawei Funding \#YBWL2010KJ013.}}

\markboth{IEEE Communications Letters, VOL.~x,
No.~x, xxxxxx~2012} {Shell \MakeLowercase{\textit{Wei et al.}}:
Upwards Scaling of Minimum-TSC Binary Signature
Sets by Sphere Decoding Algorithm}

\maketitle

\begin{abstract}
We consider upward scaling an overloaded min-TSC binary signature set and propose an optimal solution based on improved sphere decoding algorithm. Instead of previous suboptimal approach, we are guaranteed to find the optimal maximum-likelihood (ML) searching result with low complexity.

\end{abstract}

\begin{IEEEkeywords}
Binary sequences, code-division multiple-access (CDMA), signal design, total squared correlation, Welch bound.
\end{IEEEkeywords}


\section{Introduction}

\IEEEPARstart{I}{n} multiuser communication systems that follow the code-division
multiplexing paradigm, multiple signals are transmitted
simultaneously in time and frequency. Each signal, potentially
associated with a distinct user, is assigned an individual signature
(spreading code). A fundamental measure of the quality of the
code-division communication link is the total squared correlation
(TSC) \cite{1} over the set of assigned signatures. For a $K$-signal
system with signature length $L$, if the signature set is denoted by
$S=\{{\bf s}_1,{\bf s}_2,\cdots,{\bf s}_K\}$, $\|{\bf s}_i\|^2=L$,
$i=1,2,\cdots,K$, then the TSC of the signature set $S$ is defined
as the sum of the squared magnitudes of all inner products between
signatures,\\
\makebox
{\scalebox{0.95}{%
\parbox{0.5\textwidth}{\beq
TSC(S) \eqdef \sum_{i=1}^{K}\sum_{j=1}^{K}|{\bf s}_i^H {\bf s}_j|^2,
\enq}}}\\
where $(\cdot)^H$ denotes the conjugate transpose operation. For real/complex-valued signature sets, TSC is bounded from the ``Welch Bound'' $TSC(S) \ge K L \max\{K,L\}$ and signature sets that satisfy this bound with equality are called Welch-bound-equlity (WBE) sets \cite{1} \cite{2}. Algorithms and studies for the design of complex or real WBE signature sets can be found in  \cite{3}-\cite{7}.

In digital transmission systems, it is necessary to have finite-alphabet signature sets. Although the Welch bound is always achievable for real/complex-valued signature sets, this is not the case in general for binary antipodal signature sets. Hence, findings in \cite{1}-\cite{7} constitute only pertinent performance bounds for digital communication systems with digital signatures. In \cite{8Pados}, new bounds on the TSC of {\em binary} signature sets were
presented that lead to minimum-TSC optimal binary
signature set designs for almost all signature lengths and set sizes \cite{8Pados}-\cite{10}. The user capacity of
 minimum and non-minimum-TSC binary sets was identified and compared
 in \cite{12}. A procedure to find minimum-TSC binary
 signature sets with low cross-correlation spectrum was presented in
 \cite{13}.

The technical problem we consider in this manuscript is upward
scaling of an overloaded ($K>L$) min-TSC binary set. Subsets of underloaded $(K\le L)$ signatures maintain
TSC optimality and signatures can be returned and reassigned without
loss of optimality. This is not the case unfortunately, in general,
given a min-TSC overloaded set $(K,L)$ by \cite{8Pados}. Addition of a
signature, for example, may require complete redesign/reassignment
of the $(K+1,L)$ set to maintain TSC optimality. Previous suboptimal solution on this problem has been approached in \cite{Wei1} based on slowest decent method.

In this manuscript, we are relocating this problem with improved sphere decoding (SD) algorithm. The original SD algorithm was proposed in \cite{SD0} as Fincke-Pohst method, first applied to communication system of lattice code decoder in \cite{SD1} and then used for space-time decoding in \cite{SD2}. Instead of exhaustive maximum-likelihood (ML) searching, SD algorithm,  with complexity of polynomial order in the processing gain $L$ \cite{SD0}, considers only a small set of vectors within a given sphere rather than all possible transmitted signal vectors. With a proper initial searching radius setting, our proposed improved SD algorithm is guaranteed to find the optimal ML result with low complexity instead of previous suboptimal solutions.

\section{System Model}

We consider a code division multiplexing system with code length $L$
and $K\ge L$ signals (overloaded). The $K$ signals utilize a minimum
TSC optimal binary signature set $S$ designed according to
\cite{8Pados}, $S=\{{\bf s}_1,{\bf s}_2,\cdots,{\bf
s}_K\}$, ${\bf s}_i\in\{\pm 1\}^L$, $i=1,\cdots,K$. The TSC lower bound of this binary antipodal signature sets for overloaded ($K\ge L$) systems are given in Table II in  \cite{8Pados}.

When a new signal enters this system with signature ${\bf s}_{K+1}\in\{{\pm
1}\}^L$, the TSC of the $K+1$ signatures, given the signatures of the
$K$ preexisting signals, is\\
\makebox
{\scalebox{0.9}{%
\parbox{0.55\textwidth}{
\ben
TSC_{K+1|K} & = & \sum_{i=1}^{K+1}\sum_{j=1}^{K+1}|{\bf s}_i^T {\bf s}_j|^2\nonumber\\
& = & \sum_{i=1}^{K}\sum_{j=1}^{K}|{\bf s}_i^T {\bf s}_j|^2 + |{\bf s}_{K+1}^T {\bf s}_{K+1}|^2 + 2\sum_{i=1}^K|{\bf s}_{K+1}^T {\bf s}_i|^2\nonumber\\
& = & TSC_K + L^2 + 2 {\bf s}_{K+1}^T \left(\sum_{i=1}^K{\bf s}_i{\bf s}_i^T\right) {\bf s}_{K+1},\label{TSC_K+1}
\enn}}}
where $TSC_K$ denotes the TSC of the $K$ preexisting signals in the
system that utilizes a minimum TSC binary signature set. If we denote
the autocorrelation matrix of the preexisting $K$ signatures by
\beq
{\bf R}_K = \sum_{i=1}^{K}{\bf s}_i {\bf s}_i^T,
\enq
Equation (\ref{TSC_K+1}) shows that conditional minimization of $TSC_{K+1|K}$
with respect to ${\bf s}_{K+1}$ for fixed (min-TSC-valued) $TSC_K$
reduces to
\ben
{\bf s}_{K+1} = arg \min_{{\bf s}\in\{\pm 1\}^L}{\bf s}^T {\bf
R}_K {\bf s}. \label{arg_s}
\enn

The optimal ML solution by exhaustive search over all $2^L$ vectors in $\{\pm 1\}^L$ to find
the one that minimizes (\ref{arg_s}) is, of course, unacceptable
computationally even for moderate values of $L$. The work of \cite{Wei1} has proposed a suboptimal solution based on slowest decent method. In the following section, we are going to solve this problem with improved SD algorithm which gives us the optimal ML solution with low complexity.

\section{Optimal Signature Assignment}

Let the Cholesky's factorization of the autocorrelation matrix ${\bf R}_K$ yields ${\bf R}_K = {\bf U}^T {\bf U}$, where ${\bf U}$ is an upper triangular matrix. Then
\ben
{\bf s}_{K+1} = arg \min_{{\bf s}\in\{\pm 1\}^L}{\bf s}^T {\bf R}_K {\bf s} = arg \min_{{\bf s}\in\{\pm 1\}^L} ||{\bf U}\;{\bf s}||_F^2, \label{arg_s2}
\enn
where $||\cdot||_F$ denotes the Frobenius norm.

The original SD decoding algorithm \cite{SD0}-\cite{SD1}  searches through the discrete points ${\bf s}$ in the $L$-dimensional Euclidean space which make the corresponding vectors ${\bf z}\eqdef {\bf U} {\bf s}$  inside a sphere of given radius $\sqrt{C}$ centered at the origin point, i.e. $||{\bf U} {\bf s}||_F^2 = ||{\bf z}||_F^2 \le C$. This guarantees that only the points that make the corresponding vectors ${\bf z}$ within the square distance $C$ from the origin point are considered in the metric minimization.

Compared with the original SD algorithm, we have two main modifications: {\em (i)} The original SD algorithm are searching for integer points, i.e. ${\bf s}\in \mathbb{Z}^L$, while our signature searching alphabet is antipodal binary, i.e. ${\bf s}\in\{\pm 1\}^L$. Hence, the bounds to calculate each entry of the optimal signature are modified, or further tightened, according to our binary searching alphabet to make the algorithm work faster; {\em (ii)} According to the binary signature vector obtained by applying the direct sign operator on the real minimum-eigenvalue eigenvector of ${\bf R}_{K}$, denoted as ${\bf s}_{quant}^{(b)}$, we can have a very proper square distance setting as
\beq
C = {{\bf s}_{quant}^{(b)}}^T {\bf R}_K\;{\bf s}_{quant}^{(b)},\label{C}
\enq
such that the searching sphere radius is big enough to have at least one signature point fall inside, while in the meantime small enough to have only a few signature points within. As we can have this appropriate radius setting, we calculate the ${\bf s}^T {\bf R}_K {\bf s}$ metric for every candidate vector ${\bf s}$ that satisfies $||{\bf U} {\bf s}||_F^2 \le C$, such that the optimal signature assignment with minimum ${\bf s}^T {\bf R}_K {\bf s}$ metric is obtained from the improved SD algorithm directly.

Since the radius is fixed for our improved SD algorithm, the complexity uncertainty due to the radius update, which means that the radius need to be expanded if no points found in the sphere and the radius need to be reduced if too many points within the sphere, is not a question in this optimization.

Let $u_{ij}$, $i,j=1,\cdots,L$, denote the entries of matrix ${\bf U}$ in (\ref{arg_s2}). Then we are searching among ${\bf s}\in\{\pm 1\}^L$ such that
\ben
{\bf s}^T {\bf R}_K {\bf s} & = & ||{\bf U}{\bf s}||_F^2 = \sum_{i=1}^{L} \left(u_{ii}s_i + \sum_{j=i+1}^L u_{ij}s_j\right)^2\nonumber\\
& = & \sum_{i=1}^{L} q_{ii} \left(s_i + \sum_{j=i+1}^L q_{ij}s_j\right)^2\le C\label{Qs}
\enn
where $\mathbf s =[s_1,s_2,\cdots,s_L]^T$, $q_{ii}=u_{ii}^2$ for $i = 1,\cdots,L$ and $q_{ij}=u_{ij}/u_{ii}$ for $i = 1,\cdots,L$, $j = i+1, \cdots, L$.

We can start work backwards to find the entries $s_L, s_{L-1}, \cdots, s_1$ one by one.

\hspace{-\parindent}\underline{Step 1}: We begin to evaluate the last integer element $s_L$. Referring to (\ref{Qs}) and consider $q_{LL}s_L^2\le C$. For $s_L\in\{\pm 1\}$, $s_L$ can be chosen arbitrarily.

\hspace{-\parindent}\underline{Step 2}: Referring to (\ref{Qs}) again, a candidate value for $s_{L-1}$ is chosen satisfying the following
\beq
q_{LL} s_L^2 + q_{L-1,L-1} \left(s_{L-1} + q_{L-1,L}s_L\right)^2 \le C
\enq
which lead to\\
\makebox
{\scalebox{0.8}{%
\parbox{0.6\textwidth}{\ben
\left\lceil\; -\sqrt{\frac{C - q_{LL} s_L^2}{q_{L-1,L-1}}} - q_{L-1,L} s_L \;\right\rceil \le s_{L-1} \le \left\lfloor\; \sqrt{\frac{C - q_{LL} s_L^2}{q_{L-1,L-1}}} - q_{L-1,L} s_L \;\right\rfloor,
\enn}}}
where $\lceil x \rceil$ is the smallest integer greater than $x$ and $\lfloor x \rfloor$ is the greatest integer smaller than $x$. If we denote $\Delta_{L-1} = q_{L-1,L} s_L$ and $C_{L-1} = C - q_{LL} s_L^2$, and consider $s_{L-1}\in \{\pm 1\}$, the bounds for $s_{L-1}$ can be expressed as\\
\makebox
{\scalebox{0.9}{%
\parbox{0.55\textwidth}{\beq
LB_{L-1} \le s_{L-1} \le UB_{L-1},\label{sL-11}
\enq
\ben
UB_{L-1}&=&\min \left(\left\lfloor\sqrt{\frac{C_{L-1}}{q_{L-1,L-1}}}-\Delta_{L-1}\right\rfloor,1\right)\label{sL-12}\\
LB_{L-1}&=&\max \left(\left\lceil -\sqrt{\frac{C_{L-1}}{q_{L-1,L-1}}}-\Delta_{L-1}\right\rceil,-1\right).\label{sL-13}
\enn}}}
We can see that given radius $\sqrt{C}$ and the matrix ${\bf R}_K$, the bounds for $s_{L-1}$ only depends on the previous evaluated $s_L$, and is not correlated with $s_{L-2}, s_{L-3}, \cdots, s_1$.

In a similar fashion, we can proceed for $s_{L-2}$, and so on.

\hspace{-\parindent}\underline{Step L-k+1}: For the component of $s_k$, referring to (\ref{Qs}) and consider
\beq
\sum_{i=k}^{L} q_{ii} \left(s_i + \sum_{j=i+1}^L q_{ij}s_j\right)^2 \le C
\enq
will lead to\\
\makebox
{\scalebox{0.8}{%
\parbox{0.3\textwidth}{\ben
\left\lceil\; -\sqrt{\frac{1}{q_{kk}} \left(C - \sum_{i=k+1}^L q_{ii}\left(s_i + \sum_{j=i+1}^L q_{ij}s_j\right)^2\right)} - \sum_{j=k+1}^L q_{kj} s_j \;\right\rceil\nonumber\\
\le s_k \le \left\lfloor\; \sqrt{\frac{1}{q_{kk}} \left(C - \sum_{i=k+1}^L q_{ii}\left(s_i + \sum_{j=i+1}^L q_{ij}s_j\right)^2\right)} - \sum_{j=k+1}^L q_{kj} s_j \;\right\rfloor. \label{sk1}
\enn}}}\\
If we denote\\
\makebox
{\scalebox{0.9}{%
\parbox{0.55\textwidth}{\ben
\Delta_k = \sum_{j=k+1}^L q_{kj}s_j,\quad C_k = C - \sum_{i=k+1}^L q_{ii}\left(s_i + \sum_{j=i+1}^L q_{ij}s_j\right)^2
\enn}}}
and take consideration of $s_k\in \{\pm 1\}$, the bounds for $s_k$ can be expressed as
\beq
LB_{k} \le s_{k} \le UB_{k},\label{sk1}
\enq
\ben
UB_{k} & = & \min \left(\left\lfloor\sqrt{\frac{C_k}{q_{kk}}} - \Delta_k\right\rfloor, 1\right),\label{sk2}\\
LB_{k} & = & \max \left(\left\lceil -\sqrt{\frac{C_k}{q_{kk}}} - \Delta_k\right\rceil, -1\right).\label{sk3}
\enn
Note that for given radius $\sqrt{C}$ and the matrix ${\bf R}_K$, the bounds for $s_k$ only depends on the previous evaluated $s_{k+1}, s_{k+2}, \cdots, s_L$.

\hspace{-\parindent}\underline{Step L}: To evaluate the range of integer component $s_1$,  referring to (\ref{Qs}) and consider
\beq
\sum_{i=1}^{L} q_{ii} \left(s_i + \sum_{j=i+1}^L q_{ij}s_j\right)^2 \le C
\enq
will lead to\\
\makebox
{\scalebox{0.85}{%
\parbox{0.3\textwidth}{\ben
\left\lceil\; -\sqrt{\frac{1}{q_{11}} \left(C - \sum_{i=2}^L q_{ii}\left(s_i + \sum_{j=i+1}^L q_{ij}s_j\right)^2\right)} - \sum_{j=2}^L q_{1j} s_j \;\right\rceil\nonumber\\
 \le s_1 \le \left\lfloor\; \sqrt{\frac{1}{q_{11}} \left(C - \sum_{i=2}^L q_{ii}\left(s_i + \sum_{j=i+1}^L q_{ij}s_j\right)^2\right)} - \sum_{j=2}^L q_{1j} s_j \;\right\rfloor. \label{s11}
\enn}}}\\
If we denote\\
\makebox
{\scalebox{0.9}{%
\parbox{0.55\textwidth}{\ben
\Delta_1 = \sum_{j=2}^L q_{1j}s_j, \quad C_1 = C - \sum_{i=2}^L q_{ii}\left(s_i + \sum_{j=i+1}^L q_{ij}s_j\right)^2,
\enn}}}
and take consideration of $s_1\in \{\pm 1\}$, the bounds for $s_1$ can be expressed as
\beq
LB_{1} \le s_{1} \le UB_{1},\label{s11}
\enq
\ben
UB_{1} & = & \min \left(\left\lfloor\sqrt{\frac{C_1}{q_{11}}} - \Delta_1\right\rfloor, 1\right),\label{s12}\\
LB_{1} & = & \max \left(\left\lceil -\sqrt{\frac{C_1}{q_{11}}} - \Delta_1\right\rceil, -1\right).\label{s13}
\enn

In practice, $C_{L-1}$, $\cdots$, $C_1$ can be updated recursively by the following equations with initial settings $\Delta_L = 0$, $C_L = C$,
\ben
\left\{\begin{array}{llll}
\Delta_k & = & \sum_{j=k+1}^L q_{kj}s_j,\\
C_{k-1} & = & C_k - q_{kk}\left(\Delta_k + s_k\right)^2.\\
\end{array}\right.
\enn

The entries $s_L, s_{L-1}, \cdots, s_1$ are chosen as follows: for a chosen $s_L$, we can choose a candidate for $s_{L-1}$ satisfying the bounds (\ref{sL-11})-(\ref{sL-13}). If such $s_{L-1}$ does not exist, we go back and choose other $s_L$. Then search for $s_{L-1}$ that meets the bounds (\ref{sL-11})-(\ref{sL-13}) for the given $s_L$. If $s_L$ and $s_{L-1}$ are chosen, we follow the same procedure to choose $s_{L-2}$, and so on. When a set of $s_L, s_{L-1}, \cdots, s_1$ is chosen and satisfies all corresponding bounds requirements, one signature candidate vector ${\bf s}=[s_1, s_2, \cdots, s_L]^T$ is obtained. We choose the one among all candidates that gives the smallest ${\bf s}^T {\bf R}_K {\bf s}$ metric.

Note that this searching procedure will go through {\em all} candidates that satisfy ${\bf s}^T {\bf R}_K {\bf s} \le C$ and gives the one with minimum value. There is at least one candidate vector ${\bf s}_{quant}^{(b)}$ such that its entries satisfy all the bounds requirements, since that is how we set the radius value in (\ref{C}). On the other hand, the ML exhaustive binary search result ${\bf s}_{ML}^{(b)}$ that returns the minimum metric will also fall inside the search bounds, since
\beq
{{\bf s}_{ML}^{(b)}}^T {\bf R}_K {\bf s}_{ML}^{(b)} \le {{\bf s}_{quant}^{(b)}}^T {\bf R}_K\;{\bf s}_{quant}^{(b)} = C.
\enq
Hence, we are guaranteed to find the optimal ML exhaustive binary search result by the proposed improved SD algorithm.

Regarding the computational cost for the proposed improved SD based algorithm, first, eigen-decomposition needed for parameter setting of the square distance $C$ in (\ref{C}), will have complexity cost ${O(L^3)}$. In addition, for the improved SD algorithm with fixed square distance $C$, \cite{SD0} gives a complexity analysis and shows that the number of arithmetic operations is at most
\ben
\frac{1}{6}\left(2L^3 + 3L^2 -5L\right) + \frac{1}{2}\left(L^2 + 12L - 7\right)\quad\quad \quad\quad \quad \nonumber\\
 \times \left(\left(2\lfloor\sqrt{C t}\rfloor + 1\right)\left(\begin{array}{c}\lfloor 4 C t\rfloor + L - 1\\
\lfloor 4 C t\rfloor\end{array}\right) + 1\right),\label{FP-complexity}
\enn
where $t^{-1}$ is the lower bound for the entries $u_{11}^2, u_{22}^2, \cdots, u_{LL}^2$ of matrix ${\bf U}$.

Hence, the total computational cost for the proposed improved SD based algorithm still have polynomial complexity. In the literature of \cite{Wei1}, the binary signature assignment obtained on slowest descent method (SDM) has been proposed. Compared with SDM algorithm, the proposed improved SD algorithm has additional computational cost of (\ref{FP-complexity}). However, at the expense of this additional computational cost, the proposed improved SD based algorithm is guaranteed to get the optimal ML exhaustive searching result.

\section{Experimental Studies}

We consider a code-division multiplexing system $(K+1,L=16)$ for $K=16$ up to $31$ where each $(K,L)$ signature set
is optimally min-TSC designed by \cite{8Pados}. We compare the TSC performance of: (i) $TSC_{bound}$: the $(K+1, L)$ TSC lower bound of \cite{8Pados}; (ii) $TSC_{SDM}$: the $K+1$ signature added by the previous suboptimal approach of \cite{Wei1} based on slowest descent method (SDM); (iii) $TSC_{SD}$: the $K+1$ signature added by proposed improved SD algorithm in this manuscript; (iv) $TSC_{ML}$: the $K+1$ signature added by the ML exhaustive searching. For comparison purpose, we evaluate the TSC difference with the lower bound of \cite{8Pados}, i.e. $TSC_{SDM} - TSC_{bound}$, $TSC_{SD} - TSC_{bound}$, $TSC_{ML} - TSC_{bound}$, and plot in Fig. 1.

The comparison with theoretical minimum TSC bounds is very favorable. Frequently, the resulting sequence set is absolutely TSC optimal. From the detailed simulation data analysis, we find that the results of the proposed improved SD algorithm always reaches ML solution as we expected, and is superior to the previous suboptimal SDM algorithm. We demonstrate this in Table 1 with some typical simulation results to show the difference in TSC among SDM, SD, and ML methods.  The most important contribution of the proposed signature assignment in this manuscript is that it always achieves the ML results.

In Fig. 2, we repeat the simulation in a different way. Instead of always starting from an optimal min-TSC signature set $(K,L=16)$, $K = 16, 17,\cdots,31$ and only adding one signature $\mathbf s_{K+1}$ by the proposed algorithm, we just initiate once from an optimal min-TSC signature set $(K=16,L=16)$, and start upscaling signatures one-by-one consecutively by the proposed algorithm, all the way to a system of $(K+1=32,L=16)$. In other words, the initial signature set at the intermediate steps is not necessarily min-TSC. As we can see from Fig. 2, the results of the proposed improved SD algorithm always reaches ML solution again, and outperform the previous suboptimal SDM algorithm, which demonstrates the effectiveness of our proposed algorithm for any initial signature set.


\begin{center}
Table 1: TSC comparison of SDM, SD and ML\\
\begin{tabular}{|c|c|c|c|c|}
\hline
$K + 1$ & TSC with SDM & TSC with SD & TSC with ML\\
\hline
19 & 6544 & 6400 & 6400\\
\hline
23 & 9104 & 9088 & 9088\\
\hline
27 & 12304 & 12288 & 12288\\
\hline
\end{tabular}
\end{center}

\begin{figure}[hbt]
\includegraphics[width=3.9in]{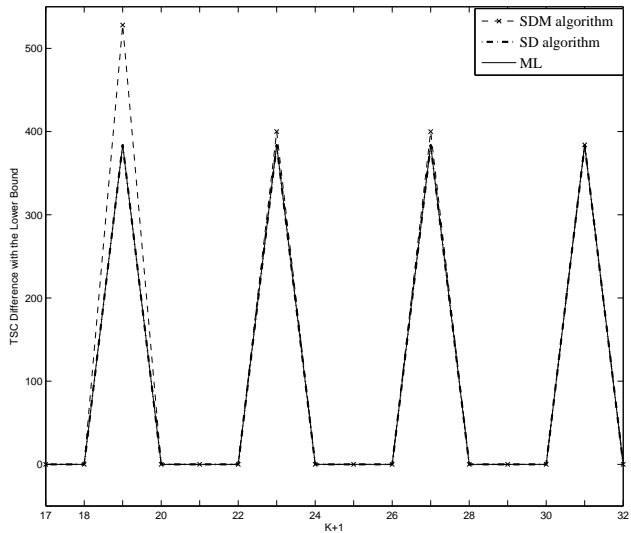}
\caption{TSC Difference with the lower bound by starting from
optimal min-TSC set at all $K=16, 17,\cdots,31$}
\end{figure}

\begin{center}
\begin{figure}[hbt]
\includegraphics[width=3.6in]{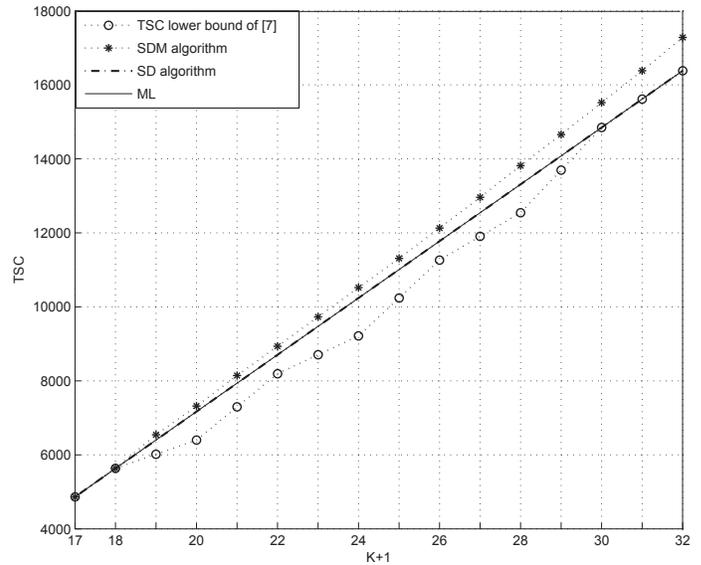}
\caption{TSC by starting from optimal min-TSC set only at $K=16$}
\end{figure}
\end{center}


\begin{thebibliography}{1}


\bibitem{1}
L. R. Welch, ``Lower bounds on the maximum cross correlation of
signals,'' {\em IEEE Trans. Inform. Theory}, vol. IT-20, pp.
397-399, May 1974.

\bibitem{2}
M. Rupf and J. L. Massey, ``Optimum sequence multisets for
synchronous code-division multiple-access channels,'' {\em IEEE
Trans. Inform. Theory}, vol. 40, pp. 1261-1266, July 1994.

\bibitem{3}
P. Viswanath, V. Anantharam, and D. N. C. Tse, ``Optimal sequences,
power control, and user capacity of synchronous CDMA systems with
linear MMSE multiuser receivers,'' {\em IEEE Trans. Inform. Theory},
vol. 45, pp. 1968-1983, Sept. 1999.


\bibitem{4}
J. Luo, S. Ulukus, and A. Ephremides, ``Optimal sequences and sum
capacity of symbol asynchronous CDMA systems,'' {\em IEEE Trans.
Inform. Theory}, vol. 51, pp. 2760-2769, Aug. 2005.


\bibitem{6}
O. Popescu and C. Rose, ``Sum capacity and TSC bounds in
collaborative multibase wireless systems,'' {\em IEEE Trans. Inform.
Theory}, vol. 50, pp. 2433-2438, Oct. 2004.


\bibitem{7}
G. S. Rajappan and M. L. Honig, ``Signature sequence adaptation for
DS-CDMA with multipath,'' {\em IEEE J. Select. Areas Commun.}, vol.
20, pp. 384-395, Feb. 2002.


\bibitem{8Pados}
G. N. Karystinos and D. A. Pados, ``New bounds on the total squared
correlation and optimum design of DS-CDMA binary signature sets,''
{\em IEEE Trans. Commun.}, vol. 51, pp. 48-51, Jan. 2003.

\bibitem{9}
C. Ding, M. Golin, and T. Kl$\phi$ve, ``Meeting the Welch and
Karystinos-Pados bounds on DS-CDMA binary signature sets,'' {\em
Designs, Codes and Cryptography}, vol. 30, pp. 73-84, Aug. 2003.

\bibitem{10}
V. P. Ipatov, ``On the Karystinos-Pados bounds and optimal binary
DS-CDMA signature ensembles,'' {\em IEEE Commun. Lett.}, vol. 8, pp.
81-83, Feb. 2004.


\bibitem{12}
F. Vanhaverbeke and M. Moeneclaey, ``Binary signature sets for
increased user capacity on the downlink of CDMA Systems,'' {\em IEEE
Trans. Wireless Commun.}, vol. 5, pp. 1795-1804, July 2006.

\bibitem{13}
P. D. Papadimitriou and C. N. Georghiades, ``Code-search for optimal
TSC binary sequences with low crosscorrelation spectrum,'' in {\em Proc. IEEE MILCOM}, Boston, MA, Oct. 2003.


\bibitem{Wei1}
L. Wei, S. N. Batalama, D. A. Pados and B. Suter, ``Upward scaling of minimum-TSC binary signature sets,'' {\em IEEE Commun. Letters}, vol. 11, no. 11, pp. 889-891, Nov. 2007.

\bibitem{SD0}
U. Fincke and M. Pohst, ``Improved methods for calculating vectors of short length in a lattice, including a complexity analysis,'' {\em Math. Comput.}, vol. 44, pp. 463-471, Apr. 1985.

\bibitem{SD1}
E. Viterbo and J. Boutros, ``A universal lattice code decoder for fading channels,'' {\em IEEE Trans. Inform. Theory}, vol. 45, no. 5, pp. 1635-1642, July 1999.

\bibitem{SD2}
O. Damen, A. Chkeif, and J. C. Belfiore, ''Lattice code decoder for space-time codes,'' {\em IEEE Commun. Letters}, vol. 4, no. 5, pp. 161-163, May 2000.

%


\end{thebibliography}
\end{document}